\documentclass[aps,pre,twocolumn,groupedaddress]{revtex4-1}

\usepackage{graphicx}% Include figure files
\usepackage{dcolumn}% Align table columns on decimal point
\usepackage{bm}% bold math
\usepackage[dvips]{color}

\def\mrm{\mathrm}

\def\etal{{\it et al. }}

\def\mrm{\mathrm}

\def\round{\partial}

\def\vep{\varepsilon}

\def\deli0{\delta_{\sigma_i 0}}
\def\delj0{\delta_{\sigma_j 0}}

\def\n+{_{n+1}}

\def \Espi{E_\mrm{spi}}
\def \bspi{\beta_\mrm{spi}}

\begin{document}

%%%%% beginning of the main doccument %%%%%%%%%%%%%%%%%%%%%%%%%%%%%%%

\preprint{APS/123-QED}

\title{
Evaporation/condensation transition 
of the two dimensional Potts model \\
in microcanonical ensemble
}% Force line breaks with \\

\author{Tomoaki Nogawa}
\email{nogawa@serow.t.u-tokyo.ac.jp}
\affiliation{%
Department of Applied Physics, 
The University of Tokyo, Hongo, Bunkyo-ku, Tokyo 113-8656, Japan
% This line break forced with \textbackslash\textbackslash
}%

\author{Hiroshi Watanabe}
% \email{watanabe@cc.u-tokyo.ac.jp}
\affiliation{%
Institute for Solid State Physics, The University of Tokyo, Kashiwanoha 5-1-5, Kashiwa,
Chiba 277-8581, Japan
}%

\author{Nobuyasu Ito}
% \email{ito@ap.t.u-tokyo.ac.jp}
\affiliation{%
Department of Applied Physics, 
The University of Tokyo, Hongo, Bunkyo-ku, Tokyo 113-8656, Japan
% This line break forced with \textbackslash\textbackslash
}%

\begin{abstract}
Evaporation/condensation transition of the Potts model on square lattice 
is numerically investigated by the Wang-Landau sampling method. 
Intrinsically system size dependent discrete transition 
between supersaturation state and phase-separation state 
is observed in the microcanonical ensemble by changing constrained internal energy. 
We calculate the microcanonical temperature, as a derivative of microcanonical entropy, 
and condensation ratio, and perform a finite size scaling of them 
to indicate clear tendency of numerical data to converge to the infinite size limit 
predicted by phenomenological theory for the isotherm lattice gas model. 
\end{abstract}

% PACS, the Physics and Astronomy
                             % Classification Scheme.
%\keywords{Suggested keywords}%Use showkeys class option if keyword
                              %display desired
%Use showkeys class option if keyword
                              %display desired
\maketitle

\section{Introduction}

First order transition 
as a state transformation of substances 
is observed in our everyday life, 
such as evaporation/condensation and melting/freezing. 
The thermodynamic mechanism of such transition is very clear; 
a phase transition occurs when the state of the lowest free energy 
alternates from one thermodynamic state to another 
by changing a certain environmental parameter. 
But dynamics of the transition is rather complicated 
and our knowledge cannot be said to be sufficient. 
Since the initial and final state is completely different unlike a second order transition, 
nucleation and large scale domain growth, 
i.e., invasion of the metastable state by the most stable state, 
is observed in the vicinity of the transition point, 
which includes various mechanisms depending on the spatio-temporal scale 
\cite{Rikvold94, Novotny02, Berg08}. 
This is essentially nonequilibrium phenomena 
and the dynamics is difficult to understand by relating it to the well-understood 
equilibrium state near the transition point. 
There is a droplet formation phenomena, however, that can be discussed 
in an equilibrium framework as mentioned in the following. 
It can be a good starting point to understand first order transition dynamics.

About first order transition, 
we usually imagine discontinuous transformation with hysteresis 
controlled by intensive variable such as temperature, pressure and magnetic field. 
But once we constrain one conjugate extensive variable, 
such as internal energy, particle density and magnetization, 
and take it as a control variable, 
a coexisting phase is inserted in the phase diagram between two homogeneous phases 
and the transitions becomes continuous; 
the volume fraction of one phase continuously changes 
from zero to unity between the two edges of the coexisting phase. 
It is pointed out, however, that a droplet, 
which is a condensed domain of the minority state in phase separation, 
is not stable unless its volume is larger than a certain threshold 
and therefore there is a discontinuous droplet condensation transition 
\cite{Biskup02, Binder03}. 
This is basically due to that the surface free energy of the droplet 
is not negligible in a finite size system. 
Therefore the threshold volume of $d$-dimensional droplet depends on the system size, 
proportional to $L^{d^2/(d+1)}$, where $L$ is a linear dimension 
of the system \cite{Binder80}. 
Although this threshold can be neglected in comparison 
with the whole system volume $L^d$ in an infinite size system, 
it actually diverges with $L$. 
Consequently, the infinite size limit of the droplet condensation transition 
is well-defined if we observe the phenomena with proper size scale. 
It should be noted that the equilibrium droplet condensation in a large size system
always contain single droplet in contrast with nonequilibrium transition 
where multiple droplets appear. 
The supersaturation state, where condensation is avoided due to small volume 
of minority state, is related to the metastable state 
under fixed intensive parameter condition.

Biskup \etal \cite{Biskup02} and Binder \cite{Binder03} 
made quantitative analysis of the equilibrium droplet condensation transition 
which is supported by a number of numerical studies 
of Lenard-Jones particles \cite{MacDowell04, MacDowell06, Schrader09} 
and lattice gas \cite{Neuhaus03, Nussbaumer06, Martinos07, Nussbaumer10} 
where particle density is constrained at given temperature. 
In this paper we consider a simpler situation 
where relevant extensive variable is only internal energy 
(magnetization is not taken account of) 
and treat the equilibrium state with constrained energy, i.e., microcanonical ensemble 
\cite{Gross96, Janke98}. 
By large scale numerical simulations of the two dimensional Potts model 
with the Wang-Landau sampling, 
we try to determine the large size limit of the droplet condensation transition, 
which has been rather difficult in small size systems \cite{Neuhaus03}.

%%%%%%%%%%%%%%%%%%%%%%%%%%%%%%%%%%%%%%%%%%%%%%%%%%%%%%%%%%%%%%%%%
\begin{figure}[t]
% \hspace{0.1cm}{\bf (a)}\hspace{7.05cm}{\bf (b)}\\ \vspace{-1.2cm}
\begin{center}
\includegraphics[trim=-10 -10 -10 -10,scale=0.210,clip]{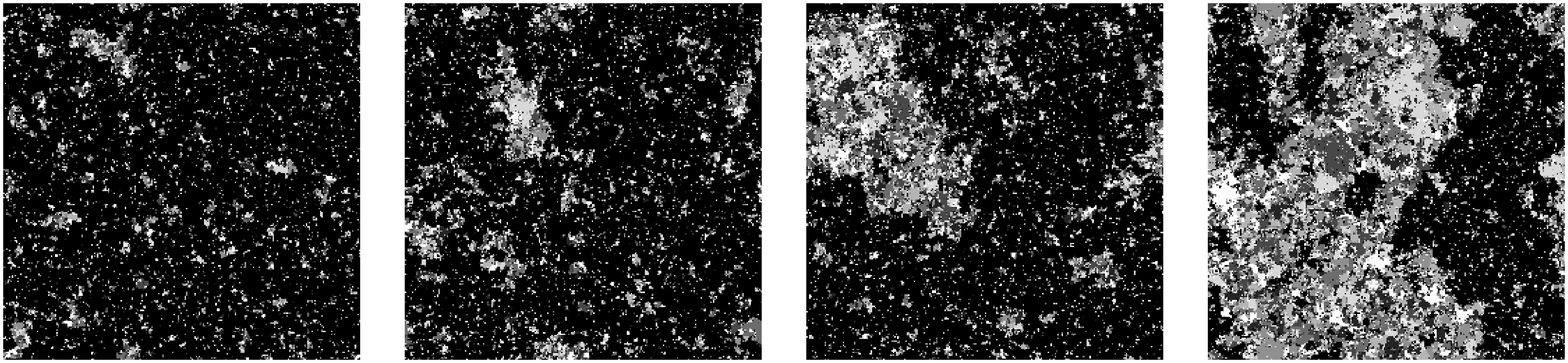}
\includegraphics[trim=-10 -10 -10 -10,scale=0.210,clip]{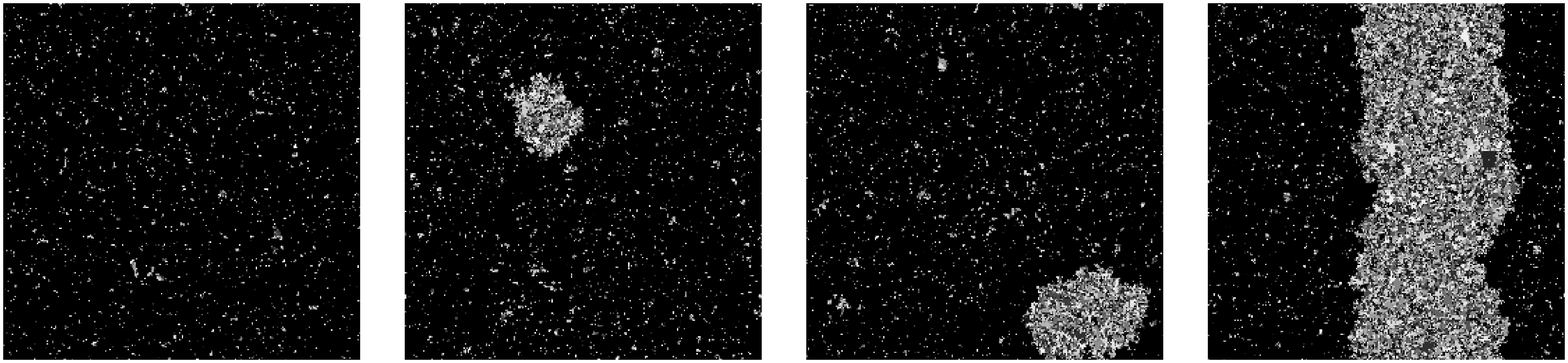}
\end{center}
\vspace{-5mm}
\caption{\label{fig:snapshot}
Typical spin configuration of the system with $q=8$ (top) and $q=21$ (bottom) 
for $L=256$. 
The different spin states are denoted by different gray scales. 
The energy is 
$E/N=0.3750, 0.4500, 0.5250$ and $0.7125$ (top) 
and 
$E/N=0.1500, 0.2250, 0.2625$ and $0.6750$ (bottom) 
from the left to the right. 
The black region is occupied by spins of the most major state 
and dappled gray region is assemblage of tiny domains 
with various states.
}
\end{figure}
%%%%%%%%%%%%%%%%%%%%%%%%%%%%%%%%%%%%%%%%%%%%%%%%%%%%%%%%%%%%%%%%%

\section{Model}

We investigate the ferromagnetic $q$-state Potts model \cite{Wu82} 
on $L \times L$ square lattice with a periodic boundary condition. 
The interaction energy for the nearest neighbor pairs is written as 
\begin{equation}
E = \sum_{\langle i, j \rangle \in \mrm{n.n.} } ( 1 - \delta_{\sigma_i \sigma_j} ), 
\end{equation}
where $\delta_{\alpha \beta}$ means Kronecker's delta 
and the spin variable $\sigma_i$ takes integer value $1, 2, \cdots, q$. 
This model in canonical ensemble exhibits paramagnetic-ferromagnetic transition 
at $\beta = \beta_c \equiv \ln( 1 + 1/\sqrt{q})$ 
where $\beta=1/k_B T$ is the inverse temperature. 
The transition is of second order for $q \le 4$ 
and of first order for $q > 4$. 
When $q$ is not sufficiently larger than 4, 
the correlation length of fluctuation is considerably large around the transition point 
and critical like fluctuation is observed to some extent.
It is better to choose large $q$ to investigate 
the pure nature of a first order transition. 
But the amount of computation for equilibration 
becomes larger with increasing $q$.

\section{Method}

We perform the Wang-Landau sampling simulations \cite{Wang01, Landau04}, 
which yields the density(number) of states $g(E)$, 
as a result of learning process to realize a flat energy histogram. 
The density of states enables us to calculate 
the Helmholtz's free energy in canonical ensemble as 
$F(\beta) = \beta^{-1} \ln \left[ \sum_E g(E) e^{-\beta E} \right]$, 
and its derivatives, i.e., mean energy and specific heat.
The energy with maximum or minimum realization probability is given 
as a solution of $\round [g(E)e^{-\beta E}]/ \round E=0$, 
i.e., $\round \ln g(E) /\round E = \beta$. 
%% Here $g(E)$ is regaded as entropy for given energy $E$. 

On the other hand, in microcanonical ensemble for given internal energy, 
fundamental thermodynamic function is microcanonical entropy $S(E) = k_B \ln g(E)$ 
and (inverse) microcanonical temperature is a quantity to be observed, 
which is defined as a response to energy perturbation as 
\begin{equation}
\beta(E) = \frac{1}{k_B} \frac{\round S(E)}{\round E} 
= \frac{\round}{\round E} \ln g(E) .
\end{equation}
This is equivalent to the extremal condition for the free energy in canonical ensemble.
% In the following, we set the Boltzmann's constant $k_B=1$. 
The $E$-dependence of $\beta$ in microcanonical ensemble 
is not exactly related to the $\beta$ dependence 
of {\it expectation} value of $E$ in canonical ensemble except in the thermodynamic limit 
but is exactly related to the $\beta$-dependence of the {\it extremal} value of $E$ 
even in the finite size system.

%%%%%%%%%%%%%%%%%%%%%%%%%%%%%%%%%%%%%%%%%%%%%%%%%%%%%%%%%%%%%%%%%
\begin{figure}[t]
% \hspace{0.1cm}{\bf (a)}\hspace{7.05cm}{\bf (b)}\\ \vspace{-1.2cm}
\begin{center}
\includegraphics[trim=0 580 0 50,scale=0.40,clip]{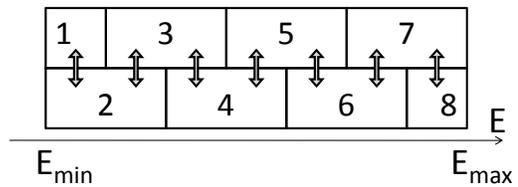}
\end{center}
\vspace{-5mm}
\caption{\label{fig:exchange}
Schematic diagram of dividing energy region to 8 threads 
and spin configuration exchange. 
}
\end{figure}
%%%%%%%%%%%%%%%%%%%%%%%%%%%%%%%%%%%%%%%%%%%%%%%%%%%%%%%%%%%%%%%%%

%%%%%%%%%%%%%%%%%%%%%%%%%%%%%%%%%%%%%%%%%%%%%%%%%%%%%%%%%%%%%%%%%
\begin{figure}[t]
% \hspace{0.1cm}{\bf (a)}\hspace{7.05cm}{\bf (b)}\\ \vspace{-1.2cm}
\begin{center}
\includegraphics[trim=0 40 40 0,scale=0.30,clip]{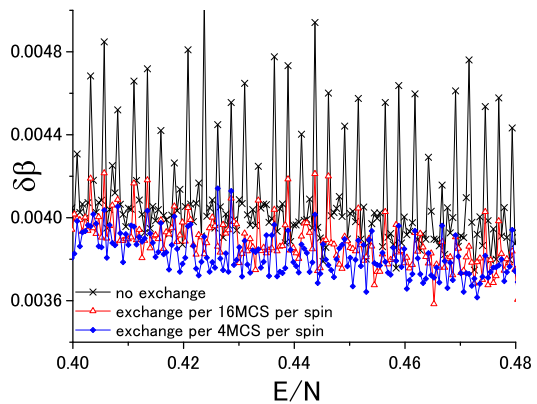}
\end{center}
\vspace{-5mm}
\caption{\label{fig:sample_dev}
(color online) 
The standard deviation of microcanonical temperature 
among samples with different realization of random number. 
The data without exchange and 4 and 16 exchanges 
in every Monte-Carlo steps per spin are plotted together. 
The simulation condition is $q=8$, $L=512$, 16 samples 
and the modification constant is down to $\ln g(E)=2^{-20}$.
The energy region, $0.36<E<0.56$, is distributed to 64 threads 
and each threads handle the energy points about $666 \times 2$. 
Smoothing is done by averaging for 128 energy points after calculating deviation.
For comparison, the difference between the transition temperature 
and the spinodal temperature, a characteristic scale of our interest, 
is about 0.001 for this size as shown in Fig.~\ref{fig:beta-E}(a).
}
\end{figure}
%%%%%%%%%%%%%%%%%%%%%%%%%%%%%%%%%%%%%%%%%%%%%%%%%%%%%%%%%%%%%%%%%

We perform parallel computation to treat large size system 
as done in Ref.~\cite{Landau04}. 
The energy region is divided into a number of parts 
with constant width and each part is associated to a different thread. 
The spin flip trials which make the energy of the system 
go out of the given range is always rejected. 
Since the time needed to diffuse the energy range $\Delta E$ by random walk 
is proportional to $\Delta E^2$, 
the time needed to obtain flat histogram is inversely proportional 
to the square of the number of threads. 
Although this method drastically reduces the total Monte-Carlo steps, 
we have to care about the possibility that 
a flat histogram is established in Monte-Carlo steps 
shorter than the relaxation time of the system. 
This occurs when the energy region is divided into too small parts.
Another problem of the division of energy region is that 
it possibly causes the segmentation of the phase space,  
i.e., there are spin configurations which cannot be 
visited depending on the initial condition. 

In order to enhance the relaxation and guarantees ergodicity, 
we make overlapping energy region for neighboring threads 
as illustrated in Fig.~\ref{fig:exchange}, 
where the exchange of spin configuration is allowed 
satisfying a detailed balance condition, 
$$
\frac{ W(\{X,Y\} \rightarrow \{Y,X\}) }
     { W(\{Y,X\} \rightarrow \{X,Y\}) }
=\frac{ g_i \left(E(X) \right) g_j \left(E(Y) \right) }
      { g_i \left(E(Y) \right) g_j \left(E(X) \right) }, 
$$
where $X$ and $Y$ are indices of microscopic states, 
in the same spirit with the replica exchange method \cite{Hukushima96}. 
Here, $g_i$ is a density of states calculated by $i$-th thread 
and $W(\{X,Y \} \rightarrow \{Y,X \})$ 
means the transition probability 
from a compound state: $X$ for the $i$-th thread 
and $Y$ for the $j$-th thread, to its exchanged state. 
Practically the exchange is accepted with a probability, 
$W(\{X,Y \} \rightarrow \{Y,X \}) = \min ( 1, g_i \left(E(Y) \right) g_j \left(E(X) \right) 
/ g_i \left( E(X) \right) g_j \left(E(Y) \right)  )$, 
similarly with the Metropolis method.

% In Fig.~\ref{fig:sample_dev}, we show the sample to sample deviation of $\beta(E)$ (find the simulation condition in the caption). It is clearly noticed that peaks observed for no exchange simulation is reduced by exchange. 
% The peaks exists at the seam point of divided energy regions, which the system can be pass through by frequent exchange. 
% Except around the seam points, however, reduction of amplitude is not large. 
% This is because the present system does not have large fluctuation for microcanonical state. 
% The exchanges will provide more benefit for more complicated systems. 

In order to check the efficiency of the replica exchange method,
the sample to sample deviations of temperature, 
$\delta \beta \equiv \sqrt{\overline{\beta^2} - |\overline{\beta}|^2}$, 
where $\overline{\cdots}$ means average over samples,
are shown in Fig.~\ref{fig:sample_dev}.
The conditions of simulations are described in its caption.
The peaks are observed in the system without exchange.
The amplitudes of the peaks becomes smaller as the
exchange frequency increases. The peaks exist at the
seam points, which are the edges of the divided region.
Except around the seam points, the reduction of fluctuations
are moderate. This is because the present system does
not exhibit large fluctuation for microcanonical state.
The exchange method can work more effectively for the system
with more complicated landscape in phase space.

\section{Results}

We perform simulation with $q=8$ and $21$ 
for system size $L=32-1024$ by using $4-64$ threads. 
The density of states is calculated not for all energy region 
but restricted region of interest.
The numerical data shown below are average over $2-8$ samples 
with different realizations of random number. 
In the Wang-Landau sampling 
we decrease the modification constant for $\ln g(E)$ step by step, 
$1, 2^{-1}, 2^{-2}, \cdots, 2^{-30}$ 
(except for $q=8$ with $L=1024$ and $q=21$ with $L=512$, 
where modification constant is decreased down to $2^{-25}$) 
with achieving 90 \% flatness of energy histogram for all threads in every step. 
Typical number of total MCSs is $10^6-10^7$ per spin 
in the final step of simulations. 
The replica exchange is attempted in every 16 MCSs per spin.

\subsection{temperature vs internal energy characteristics}

%%%%%%%%%%%%%%%%%%%%%%%%%%%%%%%%%%%%%%%%%%%%%%%%%%%%%%%%%%%%%%%%%
\begin{figure}[t]
\begin{flushleft}
\hspace{.3cm} {\large \bf (a)}
\end{flushleft}
\vspace{-.9cm}
\includegraphics[trim=20 40 80 20,scale=0.300,clip]{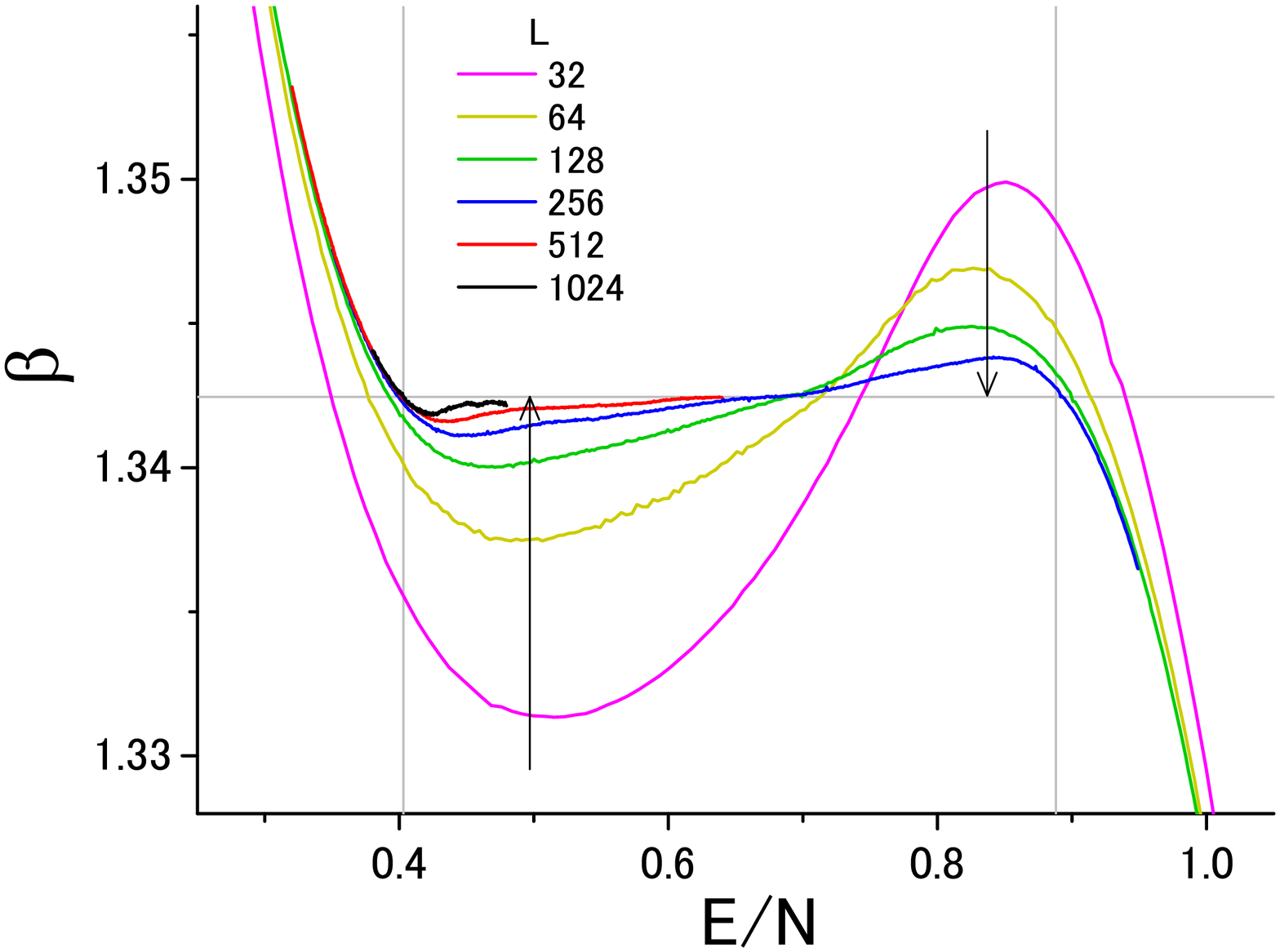}
\begin{flushleft}
\hspace{.3cm} {\large \bf (b)}
\end{flushleft}
\vspace{-.9cm}
\includegraphics[trim=20 40 80 20,scale=0.300,clip]{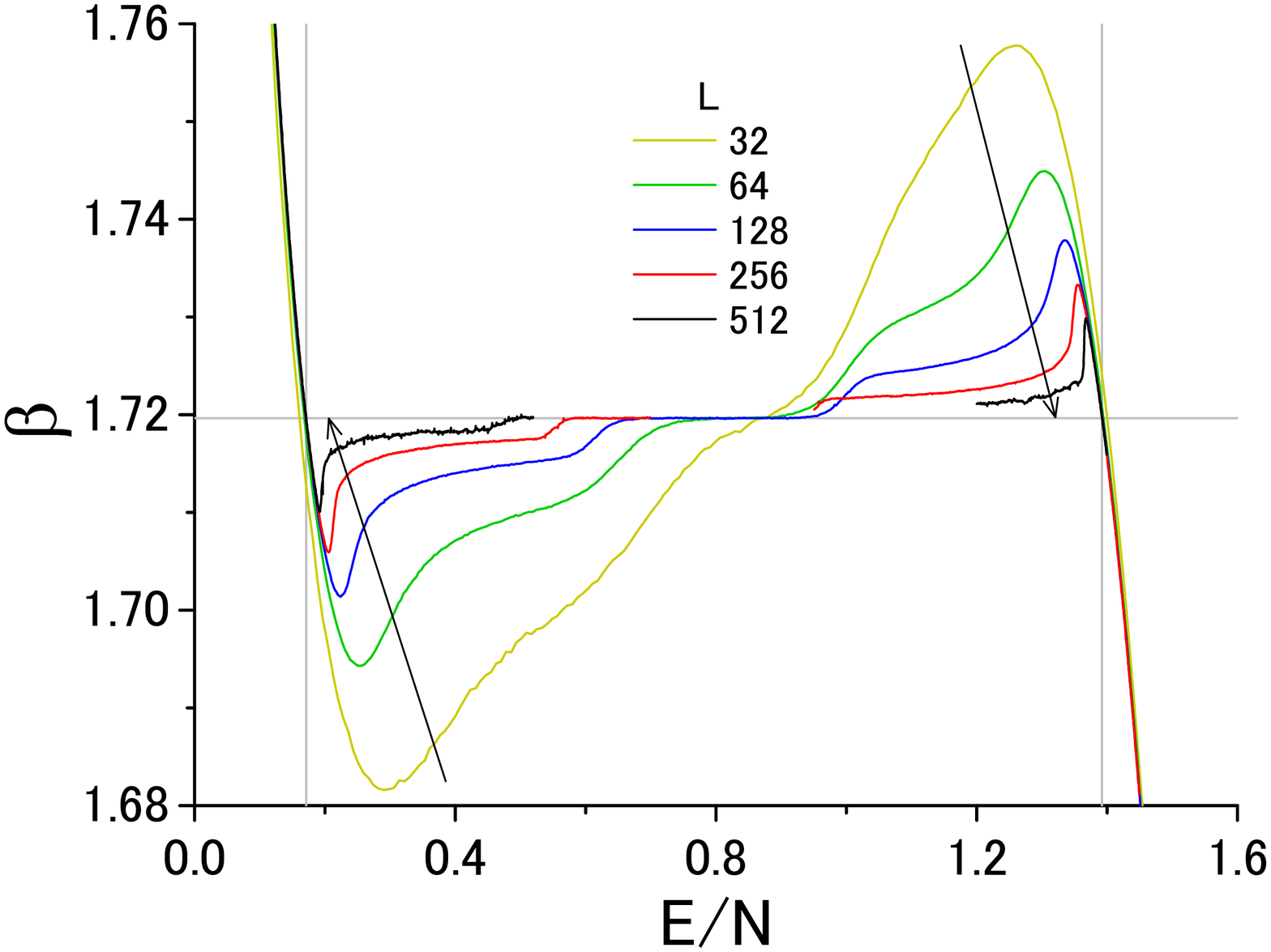}
% \end{center}
\vspace{-2mm}
\caption{\label{fig:beta-E}
(color online) 
Inverse temperature as a function of internal energy 
for (a)$q=8$ and (b)$q=21$.
Smoothing is performed by averaging over 
the range $\sqrt{N}/4$ of $E$.
The horizontal line indicates $\beta=\ln( 1 + 1/\sqrt{q})$. 
The vertical lines indicates $E/N=\vep_c^-$ (left) 
and $\vep_c^+$ (right). 
The arrows indicates the direction that $L$ becomes larger. 
}
\end{figure}
%%%%%%%%%%%%%%%%%%%%%%%%%%%%%%%%%%%%%%%%%%%%%%%%%%%%%%%%%%%%%%%%%

%%%%%%%%%%%%%%%%%%%%%%%%%%%%%%%%%%%%%%%%%%%%%%%%%%%%%%%%%%%%%%%%%
\begin{figure}[t]
\begin{flushleft}
\hspace{.8cm} {\large \bf (a)}
\end{flushleft}
\vspace{-.9cm}
\includegraphics[trim=20 0 200 20,scale=0.300,clip]{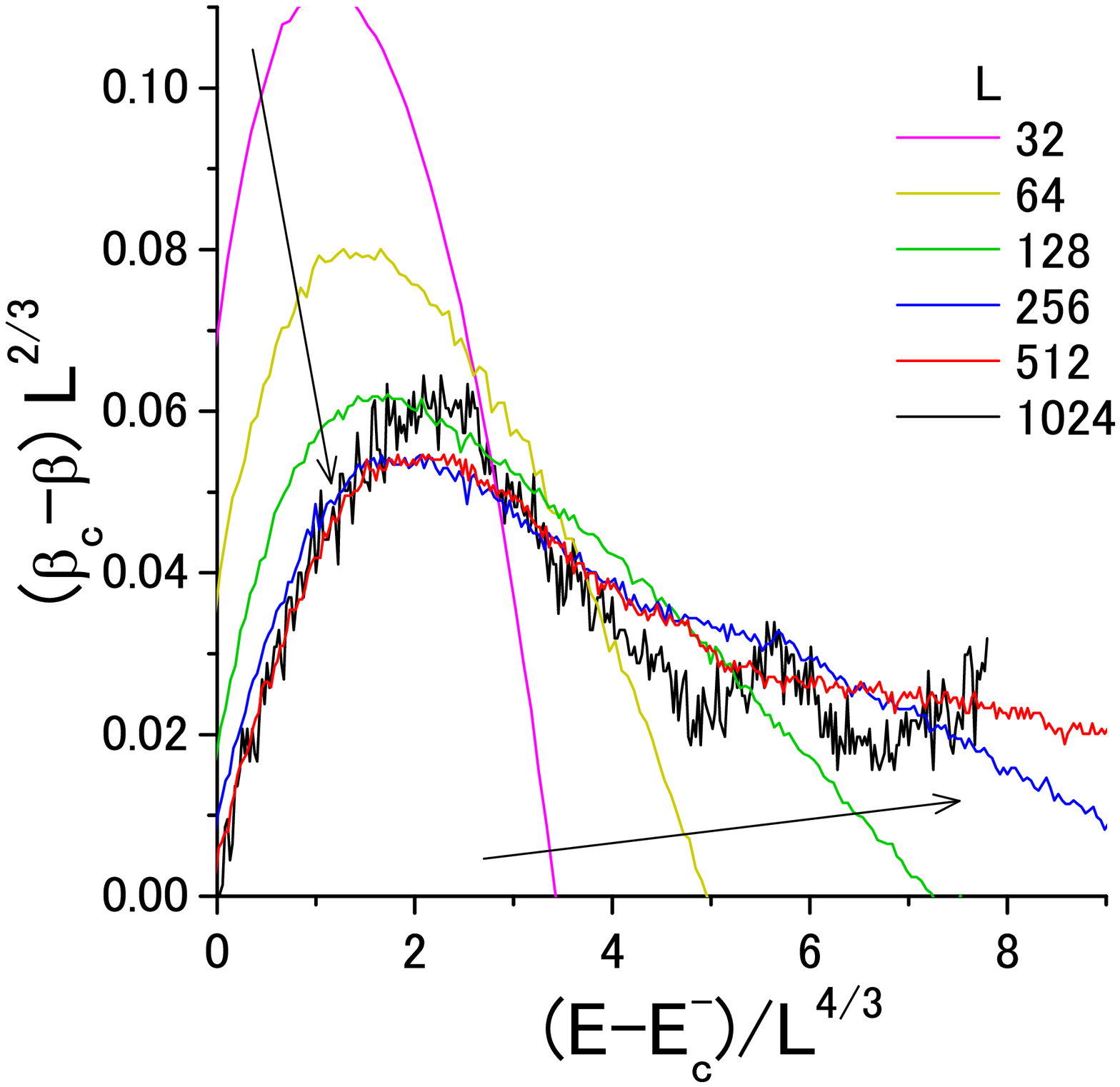}
\begin{flushleft}
\hspace{.8cm} {\large \bf (b)}
\end{flushleft}
\vspace{-.9cm}
\includegraphics[trim=20 0 200 20,scale=0.300,clip]{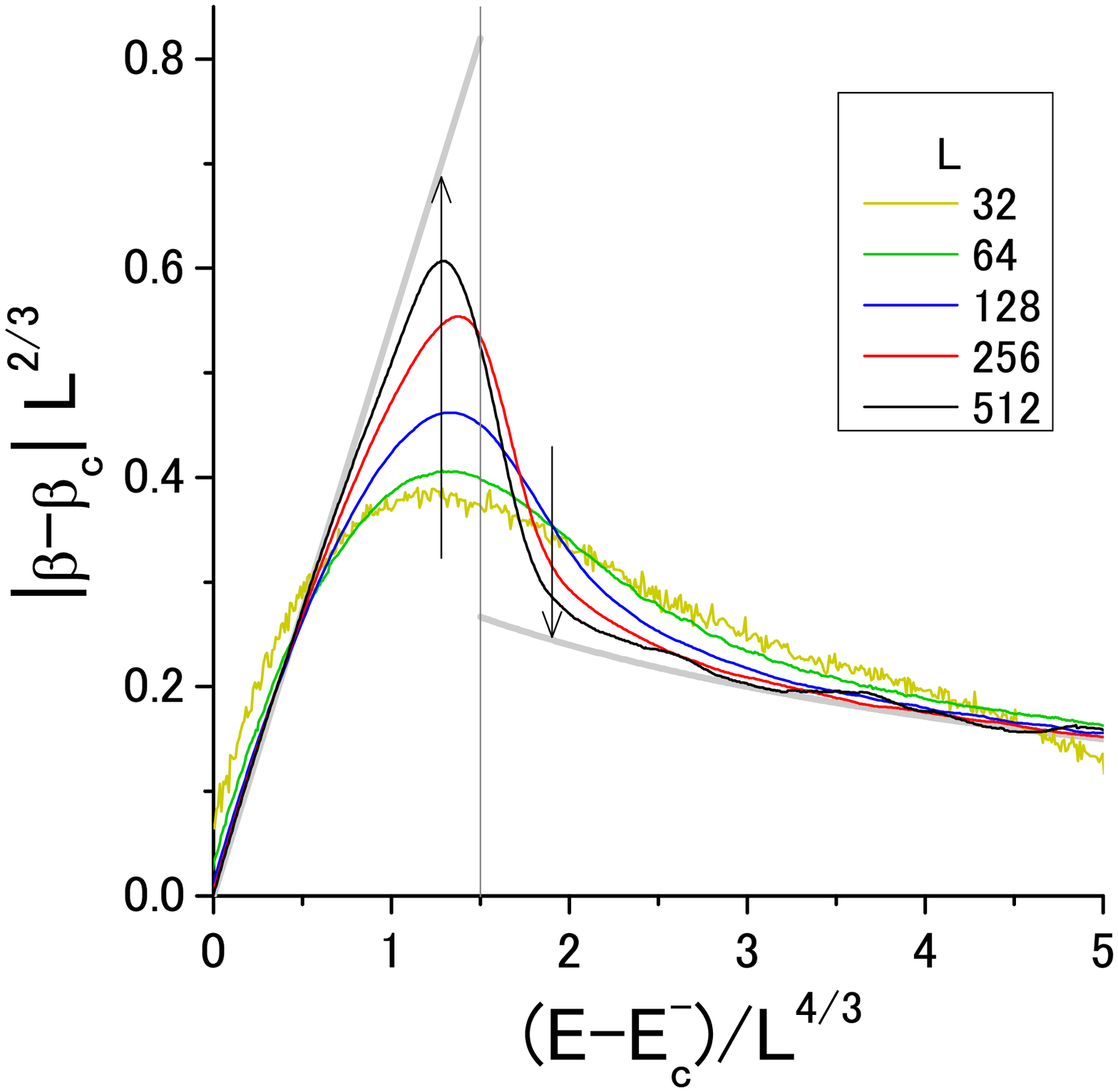}
\vspace{-2mm}
\caption{\label{fig:beta-E_scl}
(color online) 
Finite size scaling results of the excess temperature. 
(a)$q=8$ and (b)$q=21$. 
The arrows indicates the direction that $L$ becomes larger. 
The thick gray curve is a guide for eyes. 
The vertical line in the bottom panel indicates 
$E-E_c^- = 1.507 L^{4/3}$ as well as in Fig.~\ref{fig:lmd-E_scl}. 
}
\end{figure}
%%%%%%%%%%%%%%%%%%%%%%%%%%%%%%%%%%%%%%%%%%%%%%%%%%%%%%%%%%%%%%%%%

Figure~\ref{fig:beta-E} shows the inverse microcanonical temperature 
$\beta(E) = g(E+1) - g(E)$ for $q=8$ and $21$ for various system size. 
While $\beta(E)$ should be a monotonically decreasing function of $E$ 
% unlikely in the thermodynamic limit 
to make the free energy a convex function of $E$, 
it is not for the finite size system. 
In the region with positive derivative, i.e., negative specific heat, 
phase coexisting is observed as shown in Fig.~\ref{fig:snapshot}. 
This state corresponds to the free energy maximum in the canonical ensemble. 
A dip and a peak exist inside the coexisting region in the thermodynamic limit, 
$\vep_c^- < E/N <  \vep_c^+$, 
where $\vep_c^- \sim 0.403$ and $\vep_c^+ \sim 0.888$ for $q=8$ 
and $\vep_c^- \sim 0.171$ and $\vep_c^+ \sim 1.392$ for $q=21$. 
We note the bottom/top position of the dip/peak as 
$(E, \beta)$ = $(\Espi^-(L), \bspi^-(L))$ and 
$( \Espi^+(L), \bspi^+(L) )$, respectively. 
These points are regarded as the equilibrium spinodal points, 
i.e., saddle-node bifurcation points, 
where the second free energy minimum in canonical distribution annihilates 
together with one free energy maximum. 
As increasing system size, 
$\Espi^\pm/N$ approaches $\vep_c^\pm$ and 
$\beta$ approaches $\beta_c$ for all $E$ with $\vep_c^- \le E/N \le \vep_c^+$. 

For $q=21$, a plateau, where $\beta$ almost equals $\beta_c$, 
is obviously observed in the middle of coexisting region. 
In this plateau region, the two coexisting phases form strips 
parallel to the horizontal or longitudinal axis, 
while minority phase forms a droplet in the side regions 
(see Fig.~\ref{fig:snapshot}). 
With increasing system size, it is observed that 
the plateau region becomes wider 
and the step-like change of $\beta(E)$ on the edge becomes sharper 
as a consequence of droplet-strip(slab) transition \cite{Neuhaus03}. 

As increasing system size, the two spinodal points, $(\Espi^\pm(L), \bspi^\pm(L))$, 
approach $(E_c^\pm, \beta_c) \equiv (N \vep_c^\pm, \beta_c)$, respectively. 
The deviation $|\bspi^\pm(L) - \beta_c|$ and $|\Espi^\pm(L) - E_c^\pm|$ 
decreases as a power function of $L$ \cite{Bazavov08}. 
% The correction to the power-law for small $L$ becomes smaller for larger $q$. 
To show it clearly, we perform finite size scaling 
in Fig.~\ref{fig:beta-E_scl} with expecting a formula 
\begin{equation}
| \beta(E) - \beta_c | = L^{-d/(d+1)} 
F \left( \frac{|E-E_c^\pm|}{L^{d^2/(d+1)}} \right), 
\end{equation}
with a scaling function $F(\cdot)$, 
that for isotherm lattice gas model is derived in Ref.~\cite{Binder03}, 
where $\beta$ and $E$ are replaced with magnetic field and magnetization, respectively. 
For $q=21$, it is observed that the scaled curve approaches 
a large size limit; 
$|\beta-\beta_c|$ linearly increases as $(k_B \beta_c^2/C^\pm) |E-E_c^\pm|$ 
where $C^\pm$ is a specific heat, 
$- k_B \beta^2 [d^2 \ln g(E)/dE^2]^{-1}$ at $E=E_c^\pm$ 
in the evaporation phase
and discontinuously drops down when entering the condensation phase.
Note that this scaling does not aim the collapse of data to a universal curve 
as in standard finite size scaling for second order transitions, 
but it is intended to show the conversion to the large size limit 
by blowing up the transition region, which vanishes in macroscopic scale. 
(Scaling of finite size rounding may be a challenging problem.) 
On the other hand, discontinuous behavior is rarely observed 
in the system with $q=8$ even for $L=1024$. 
This is because correlation length is rather large 
and the used sample size is effectively much smaller than that for $q=21$.

\subsection{volume fraction of a droplet}

%%%%%%%%%%%%%%%%%%%%%%%%%%%%%%%%%%%%%%%%%%%%%%%%%%%%%%%%%%%%%%%%%
\begin{figure}[t]
% \hspace{0.1cm}{\bf (a)}\hspace{7.05cm}{\bf (b)}\\ \vspace{-1.2cm}
\begin{center}
\includegraphics[trim=20 0 60 20,scale=0.280,clip]{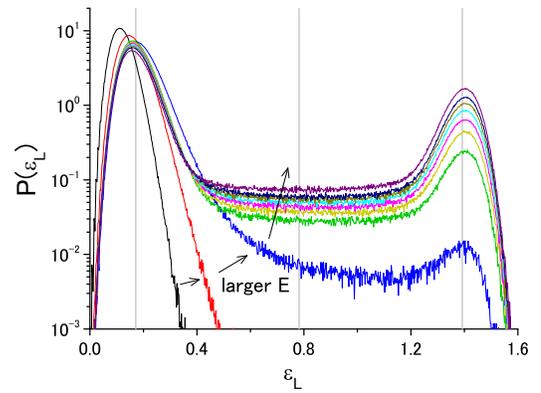}
\end{center}
\vspace{-5mm}
\caption{\label{fig:P-e}
(color online) 
Probability distribution function of the energy density of subsystems 
for $q=21$, $L=512$. 
Each curves corresponds to $E/N$=0.12-0.52 (0.02 step). 
}
\end{figure}
%%%%%%%%%%%%%%%%%%%%%%%%%%%%%%%%%%%%%%%%%%%%%%%%%%%%%%%%%%%%%%%%%

To observe the nature of droplet formation more directly, 
we estimate the ratio of condensation volume to the whole system volume. 
To this end, we divide a sample into $L$ square-shaped subsystems 
with size $\sqrt{L} \times \sqrt{L}$ 
( $\sqrt{L}$ is approximated by the closest integer if necessary). 
This resolution is fine enough to capture 
the shape of the critical droplet with size of $O(L^{d^2/(d+1)})$. 
We calculate energy per spin $\vep_L$ for each square, 
whose distribution function $P(\vep_L)$ is shown in Fig.~\ref{fig:P-e}. 
Hereafter we only show the results for $q=21$. 
Bimodal distribution is observed in the condensed regime, $\Espi^-(L) < E < \Espi^+(L)$. 
The width of peaks becomes narrower as $\sqrt{\vep_L L }/L \sim L^{-1/2}$ 
and the integral of bridge component between the two peaks 
corresponding to the perimeter of droplet 
becomes smaller with $(2 \pi L \times \sqrt{L})/N = L^{-1/2}$. 
The positions of the peaks hardly change with $E$ 
but only heights change for $\Espi^-(L) < E < \Espi^+(L)$, 
which means that the change of state in this regime 
can be described only by the change of mixing ratio of the two phases. 
We determine a subsystem with energy 
below/above $\vep_m = (\vep_c^+ + \vep_c^-)/2 = ( 1 - 1/\sqrt{q} )$ \cite{Wu82}
belongs to the ferro/para domain.

%%%%%%%%%%%%%%%%%%%%%%%%%%%%%%%%%%%%%%%%%%%%%%%%%%%%%%%%%%%%%%%%%
\begin{figure}[t]
% \hspace{0.1cm}{\bf (a)}\hspace{7.05cm}{\bf (b)}\\ \vspace{-1.2cm}
\begin{center}
\includegraphics[trim=20 0 40 20,scale=0.320,clip]{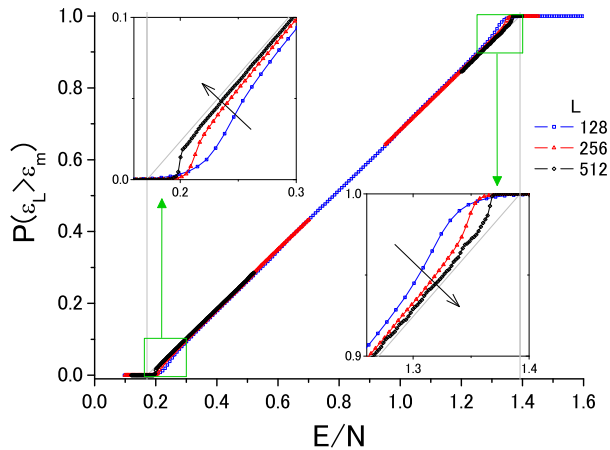}
\end{center}
\vspace{-5mm}
\caption{\label{fig:V-E}
(color online) 
Internal energy dependence of the volume fraction of the high energy domain. 
The insets are the blowups of the edges of the coexisting phase, 
where the arrows indicates the direction that $L$ becomes larger. 
}
\end{figure}
%%%%%%%%%%%%%%%%%%%%%%%%%%%%%%%%%%%%%%%%%%%%%%%%%%%%%%%%%%%%%%%%%

\begin{figure}[t]
% \hspace{0.1cm}{\bf (a)}\hspace{7.05cm}{\bf (b)}\\ \vspace{-1.2cm}
\begin{center}
\includegraphics[trim=20 0 200 20,scale=0.300,clip]{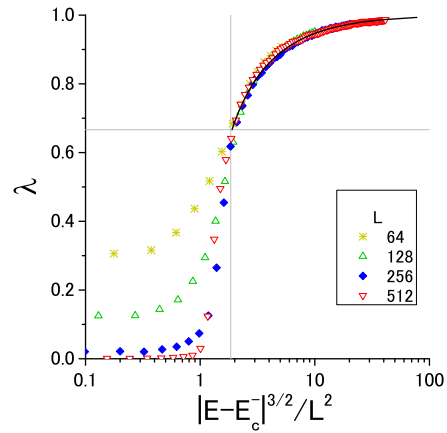}
\end{center}
\vspace{-5mm}
\caption{\label{fig:lmd-E_scl}
(color online) 
Finite size scaling result of the volume fraction of condensed droplet. 
The data for $E/N<0.40$ are used. 
The solid gray curve indicates theoretical prediction, 
$a |E-E_c^-|^{3/2}/L^2 = 1/4 \sqrt{\lambda} ( 1 - \lambda )$ for $\lambda>2/3$, 
where we set $a=0.44$. 
The vertical line indicates $E-E_c^- = 1.507 L^{4/3}$ 
as well as in Fig.~\ref{fig:beta-E_scl}.
}
\end{figure}

The volume fraction of the para phase, $P(\vep_L > \vep_m)$, 
is plotted in Fig.~\ref{fig:V-E}. 
In the thermodynamic limit, 
$P(\vep_L > \vep_m)$ equals zero for $E \le E_c^-$, unity for $E \ge E_c^+$ 
and linearly increases  in between as $(E-E_c^-)/(E_c^+-E_c^-)$.
Finite size deviation is observed on the edge of the coexisting region 
as shown in the insets of Fig.~\ref{fig:V-E}.
The fraction of condensed phase for finite size system 
is smaller than that for $L=\infty$. 
The normalized condensation ratio 
\begin{eqnarray}
\lambda \equiv \frac{P(\vep_L>\vep_m)}{(E-E_c^-)/(E_c^+-E_c^-)} \le 1
\end{eqnarray}
is expected to be a function of dimensionless variable,  
\begin{eqnarray}
\Delta = a \frac{ (E-E_c^-)^{(d+1)/d}}{ L^d} 
\\
\mrm{with} \quad 
a = \frac{ \beta_c ( \vep_+ - \vep_- )^{(d-1)/d} }{ 2 C^- \tau_W }
\end{eqnarray}
for $L \rightarrow \infty$. 
Here $\tau_W$ is an interface free energy per volume 
of an optimally shaped large Wulff droplet \cite{Wulff01}. 
The derivation of this formulae is described in Appendix \ref{ap1}. 
The behavior in the thermodynamic limit is exactly known \cite{Biskup02} as 
$\lambda = 0$ for $\Delta < \Delta_c \equiv (1/2)(3/2)^{3/2}$ and 
$1/4 \sqrt{\lambda} ( 1 - \lambda ) = \Delta$ for $\Delta > \Delta_c$ .
While $C^-$ is estimated as 5.41, $\tau_W$ is an only unknown quantity 
needed to calculate $\Delta$. 
By only assuming $\tau_W=0.40$ (then $a=0.44$), however, our numerical result 
shows good agreement with the theory as shown in Fig.~\ref{fig:lmd-E_scl}.
While $\lambda$ decreases to zero for $\Delta<\Delta_c$ with increasing $L$, 
very little finite size dependence is observed $\Delta>\Delta_c$.
The gap of $\lambda$ at $\Delta_c$ also coincides with the predicted value $2/3$.

%%%%%%%%%%%%%%%%%%%%%%%%%%%%%%%%%%%%%%%%%%%%%%%%%%%%%%%%%%%%%%%%%
%%%%%%%%%%%%%%%%%%%%%%%%%%%%%%%%%%%%%%%%%%%%%%%%%%%%%%%%%%%%%%%%%

\section{Conclusion}

We have investigated the condensation/evaporation transition 
of the Potts model in microcanonical ensemble.
Interesting property of this transition is that 
it becomes impossible to be found in the true thermodynamic limit; 
the width of supersaturation regime disappears, 
but discontinuity of scaled quantities 
becomes clear with increasing system size. 
The present numerical results with $q=21$ show good agreement with the theoretical prediction 
of the system size dependence both for temperature \cite{Binder03} 
and condensation ratio \cite{Biskup02}. 
Both of the two mean that droplet with size smaller 
than $O(L^{d^2/(d+1)})$ is unstable.

For finite size system in microcanonical ensemble, 
we observe negative (inverse) specific heat in the coexisting region 
and over/underhang of temperature, that corresponds to 
the thermodynamic spinodal point in canonical ensemble. 
The scaling behavior $| \beta_\mrm{spi} - \beta_c | \propto L^{-d/(d+1)}$ 
suggests an existence of a diverging length scale, 
$R_s(\beta) \propto | \beta - \beta_c |^{-(d+1)/d}$. 
This means that a supersaturation state at given temperature $\beta$ 
becomes unstable at a length scale above $R_s$. 
% This diverges faster than the critical nucleus size $R_c \propto (\beta - \beta_c)^{-1}$.
Although this length is related to the equilibrium spinodal point, 
it is not clear whether it also has some meanings in nonequilibrium dynamics, 
which may be an interesting open problem.

Last of all, let us note on the difficulty of the simulation of first order transitions. 
We suffered from slow relaxation for large system size in the present study 
as well as discussed in Ref.~\cite{Neuhaus03}. 
It is considered that the Wang-Landau sampling is quite efficient 
for first order transitions with discontinuity of $O(L^d)$ 
as well as some other extended ensemble methods, 
such as the multicanonical method \cite{Berg92} 
because it provides additional probability weight on coexisting states 
to bridge the divide between two distinct homogeneous states, 
such as para and ferro phases. 
However there are still discontinuous transitions in the coexisting phase, 
that is a evaporation/condensation transition with $O(L ^{d^2/(d+1)})$ discontinuity
and droplet/slab transition, \cite{Neuhaus03, Nussbaumer06}.  
This behavior is general for first order transitions. 
Since the problem is that the two states are energetically degenerate in these transitions, 
one fundamental solution to this problem may be 
to employ another argument which corresponds to the shape of domains 
for the joint density of state \cite{Landau04, Zhou06}. 
Although it requires considerable amount of computation, 
massive parallel computation makes it feasible.

\appendix

\section{finite size effect on droplet condensation ratio}
\label{ap1}

Here we derive the condensation rate of a droplet 
for a energy driven phase transition 
by translating the result for a magnetization driven transition \cite{Biskup02, Nussbaumer10}. 
We consider canonical ensemble at the bistable point $\beta = \beta_c$ 
and its energy distribution function around a peak at $E_c^-$.
(The case around the another peak $E_c^+$ is derived in the same way.  )

We note the excess energy in fluctuation beyond the peak as $\delta E = E - E_c^-$ 
and divide into two parts, $\delta E = \delta E_b + \delta E_d$, 
where $\delta E_b$ is due to small bubble excitation 
and $\delta E_d$ is due to a large droplet of disordered phase. 
If introducing condensation ratio $\lambda$ as 
$\delta E_d \equiv \lambda \delta E$, 
the rest is given by $\delta E_b = (1-\lambda) \delta E$.
The volume of the droplet, $V_d \equiv \lambda V_L$, 
can be smaller than that in the thermodynamic limit, 
$V_L = \delta E/(\vep_c^+ - \vep_c^-)$, for finite size system.
By using these quantities, 
the energy distribution function $P(\beta_c; E)$ 
is proportional to $e^{-\beta_c F}$, where 
\begin{eqnarray}
F &=& \frac{ \beta_c (\delta E_b )^2}{2 L^d C^-}
+ \tau_W V_d^{(d-1)/d} 
\\
&=& \tau_W V_L^{(d-1)/d} \left[
\Delta (1-\lambda)^2 + \lambda^{(d+1)/d}
\right] 
\\
\mrm{with} &&
\Delta \equiv \frac{\beta_c (\vep_c^+ - \vep_c^-)^{(d-1)/d}}{2C^- \tau_W}
\frac{(\delta E)^{(d+1)/d}}{L^d}.
\end{eqnarray}
The first term indicates the fluctuation without phase coexistence 
characterized by specific heat $C^-$ 
and the second term indicates the surface free energy of a large droplet.

The nature of the fluctuation depends on the dimensionless parameter $\Delta$ 
which is related to the excess energy.
For given $\Delta$, free energy $F$ is minimized at 
\begin{equation}
\lambda = 0 \quad \mrm{for} \quad \Delta \le \Delta_c \equiv (1/2)(3/2)^{3/2}, 
\end{equation}
which means supersaturation regime, 
while $\lambda$ for minimum $F$ is given by a solution of 
\begin{equation}
1/4 \sqrt{\lambda} ( 1 - \lambda ) = \Delta
\quad \mrm{for} \quad \Delta \ge \Delta_c,
\end{equation}
which means condensation regime.
The two states with $\lambda=0$ and $\lambda \equiv \lambda_c = 2/3$ 
is equivalently stable at $\Delta_c$. 

This discontinuous change at $\Delta_c$ is directly observed 
as a internal energy driven transition in microcanonical ensemble.

%\acknowledgments
    
\vspace{.20cm}
\begin{center}
{\bf ACKNOWLEDGMENTS}
\end{center}

\vspace{-.20cm}

This work was partly supported by Award No. KUK-I1-005-04 made by King Abdullah University of Science and Technology (KAUST).

% \bibliography{d:/home/mydoc/study/mybib.bib}

% \input{jja_gs13.bbl}

% \begin{thebibliography}{99} 
% \input{jja_gs13.bbl}
% \end{thebibliography}

%merlin.mbs apsrev4-1.bst 2010-07-25 4.21a (PWD, AO, DPC) hacked
%Control: key (0)
%Control: author (8) initials jnrlst
%Control: editor formatted (1) identically to author
%Control: production of article title (-1) disabled
%Control: page (0) single
%Control: year (1) truncated
%Control: production of eprint (0) enabled
%

\end{document}